# First Principles and Machine Learning Identify Key Pairing Strength Factors of Cuprate Superconductors

**Xinyu He[1], Ning Chen[1], Jingpei Chen[1], Xuezhou Wang[1*], Yang Li[2*]**

**[1]School of Materials Science and Engineering, University of Science and Technology Beijing, Beijing 100083, China**

**2 Department of Engineering Science and Materials，University of Puerto Rico，Mayaguez，Puerto Rico 00681-9000，USA**

***Corresponding author：nchen@sina.com； yang.li@upr.edu**

## Abstract

By using band structure calculations of quantum mechanical theory, some important peaks of DoS (Density of States) were obtained and classified based on crystal structure laws of cuprate superconductivity. In particular, the orbital interactions of the in-plane and out-of-plane ions of the copper-oxygen plane were investigated. The position, half-width, and height of DOS peak features were collected for all 35 typical curate systems which have critical temperature maximum data from the works of literature. By training test of 7 common machine learning algorithms, the relationship between the Tc maximum values and these orbital interaction parameters were mined. It was found that the key features of the orbital interaction affecting the *T*c maximum were not only the flat band but also a new interaction between core orbitals in a deeper energy band position.

## Introduction

The electron pairing mechanism of high-temperature superconductivity has been one of the most challenges of condensed matter physics for almost four decays[1] as a result of a number of complex and deeply hidden factors influencing on the critical temperature ($T_c$) of superconductivity, where $T_c$ represents the stiffness of superconductivity which combines with both electronic pairing factors and carrier concentration or doping factors [2]. By means of experimental studies on $T_c$ variations with environment, composition and structure of materials, these above two kinds of influence factors had been found[3] and some complex relationships were also understood through machine learning (ML) studied on a large data of 12,000 $T_c$ values.[4,5]

However, the governing strength of pairing factors remains unknown because of the following two technical problems. The first one is that the $T_c$ Raw data's weights distribute mainly on doping factors but less on pairing ones. For example, nearly 6000 experimental data of $T_c$ values cover mainly 35 of typical systems with only one *T*c maximum value for each system, or only 6% data weight of pairing factor, so that the predicted ML model was trained out not a new system with a higher $T_c$ maximum or a stronger pairing factor, but only a new one with suitable doping in fact. [6,7] As the $T_c$ maximum is different in each system for a similar optimal doping of copper oxides [8], the pairing strength of one system should be replaced by its *T*c maximum values for ML training test in order to search for pairing factors.

On the other hand, the second problem is that these pairing factors were mainly observed from features or attributions of atoms or ions rather than explicit orbitals or electronic interactions[9]. Although electronic structures could be easily obtained by the first principal approach of quantum physics, it is hard to tackle with both these doped models' complex constructions and these burden computations for thousands of doping systems. Even if this problem were solved, the limited data set problem still remains due to a much low weight ratio on the pairing factor. Therefore, changing target data set of the *T*c maximum as well as training by key orbital features is the only way to solve the existing technical problems.

## Method

For cuprates, there are three crystal categories: the layer structure for Hg, Tl, Pb and Bi families, the 123 structure and the 214 structure. But even for one structure, there is also a lot of complex orbital interaction features to be considered. Here we should provide a simpler method which is based on orbital peaks of interacting bands from Density-of-State (DoS) diagrams obtained by band structure calculations of First principles, in which one peak represents only one band, ignoring its orbital peaks in details. In addition, its position, half-width and height, quite different from its atomic orbital, are chosen as their peaks' features to discuss the interactions between different orbitals.

As result of lacks of $T_c$ maximum data values, the number of influencing features for ML training had to be limited by introducing domain knowledge, so that crystal structure laws of cuprate superconducting are quite important to help us for cutting off orbital feature's candidates. In this paper, we especially analysis a relatively larger energy range of energy band to cover most of important *s*, *p* and *d* peaks of DoS. [10] Firstly, the copper-oxygen plane is a main superconducting "highway" for high temperature superconductivity, oxygen ion's 2*p* and copper ion's 3*d* orbitals need to be focused on, which were thought to be a common feature of cuprate superconductors. Secondly, as a difference of key crystal structure of each cuprate system mainly comes from the nearest neighboring cation beside the copper-oxygen plane,[11] so that the second outer saturated orbitals of them had to be taken into account. Similar reason for the second outer saturated 2*s* orbital of oxygen ion since its orbital is similar in energy and space. Therefore, much deeper range of orbitals up to -30e*V* below Fermi level were studied in this work, which is quite different from in other works (Up to -10e*V*), in order to emphasize on all of these important orbital interactions around the copper-oxygen plane.

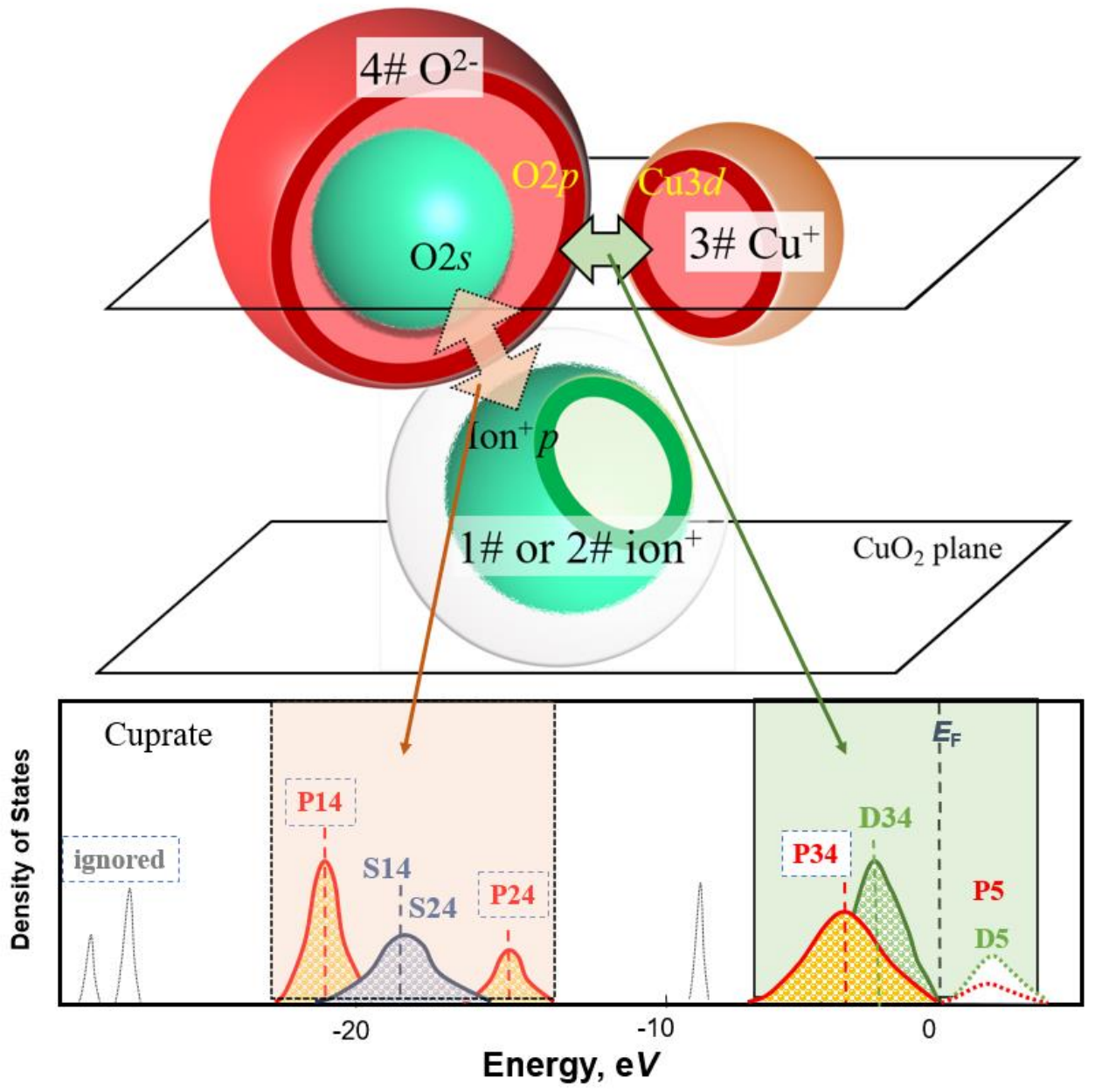

**Figure 1: Schematic diagram of Key four ions and two energy-bands for the coper-oxygen plane of cuprate and the classification of some important peaks in DoS; 1#, 2# for neighboring cations with deeper and shallower orbital levels copared to $E_F$ respectively, 3# for copper ions and 4# for oxygen ions**

Figure 1 show that a schematic diagram of some important orbitals in crystal structure, as well as their classification of orbital peaks in DoS corresponding to a typical part of cuprate crystal structure. The calculation procedure and parameters are detailed in reference[7]. Especially, we focused on 1#~4# four types of ions in the near-neighbor position outside the plane (1# Ions occupies a relatively lower *p* orbital quite far away from Fermi level, and 2# Ions occupies for a relatively higher one close to Fermi level) and in the copper-oxygen plane (3# is Cooper Ions and 4# is Oxygen Ions). Here, the ionic classification is defined for the propose of studying the interaction between some important orbitals in two energy-bands, in which it is believed that the electron orbit interaction in more detail contrary to a strength and a type of chemical bond. Therefore, total four key ions coupled with two different energy-bands composed of those 8 important peaks selected from DoS of the first principal calculations, therefore, there are total 21 basic features (position, half-width and height) of orbital peak collected for ML training. Actually, it should be noticed that the most important selection principle come from not only basic structure laws of cuprate systems but also the simplification of some important orbital coupling parameters.

The standardized identification of the position of the orbital peaks is relatively important, which is easier to identify relatively independent peaks but it is difficult to identify peaks with orbital coupling phenomena so that some special definition shou be taken. Here S, P, and D represent orbitals derived from the *s*, *p*, and *d* orbitals, so that S14 or S24 actually represent the 2*s* orbitals of the oxygen ion influenced by 1# or 2# neighboring cations, P14 or P24 represent the *p* orbitals of the 1# or 2# neighboring cations respectively as they are influenced by 2*s* of 4# oxygen cations, and also D43 represents the 3*d* orbitals of the 3# copper ion, which is influenced by the 2*p* orbitals of the 4# oxygen ion, as well as also P34 represents the 2*p* orbitals of the 4# oxygen ion, which is influenced by the 2p orbitals of the 3# copper ion. Finally, P5 and D5 represent the *p* and *d* orbital peaks belonging to these vacant states of conductive band. Respectively, sub-title _L, _W, _H represent the energy location of peak, the half-wide of peak and the height of peak for all those 8 peaks, therefore, all of 35 systems' the *Tc* maximum with total 21 parameters were set as raw data set used for machine learning, as listed in **Appendix A**.

## Results

Based on the data in **Appendix A**, hundreds of training tests were done by 7 commonly used machine learning algorithms, and coupled with different training routes and test sample selection. And the predicted model results obtained from the best results of all different training tests for different algorithms.

Here, it is quite important for us to discuss the SHAP (Shapley Additive ExPlanations) value analysis, by which we can understand the weight distribution of each peak feature on the $T_c$ maximum value. As shown in Figure 2a, the vertical coordinate represents 21 different orbital features, the horizontal coordinates are their corresponding SHAP values, and the red and blue indicate positive and negative correlations with $T_c$ respectively. Figure 2a shows the SHAP analysis of an elastic network regression model, which is quite similar to the average value of total 7 methods. Among 21 compared features, it shows that the stronger ones come from P14_L, S14_L, P24_L, D43_W and positively correlated with $T_c$ maximum, while P14_L is negatively correlated with $T_c$ maximum, and S14_L, P24_L, D43_W, positively correlated with $T_c$ maximum. Figure 2b also shows a schematic diagram of the training test results. It shows that the deeper the P14 orbital position, the larger the $T_c$ maximum. And some other order of peaks with respect to $T_c$ maximum.

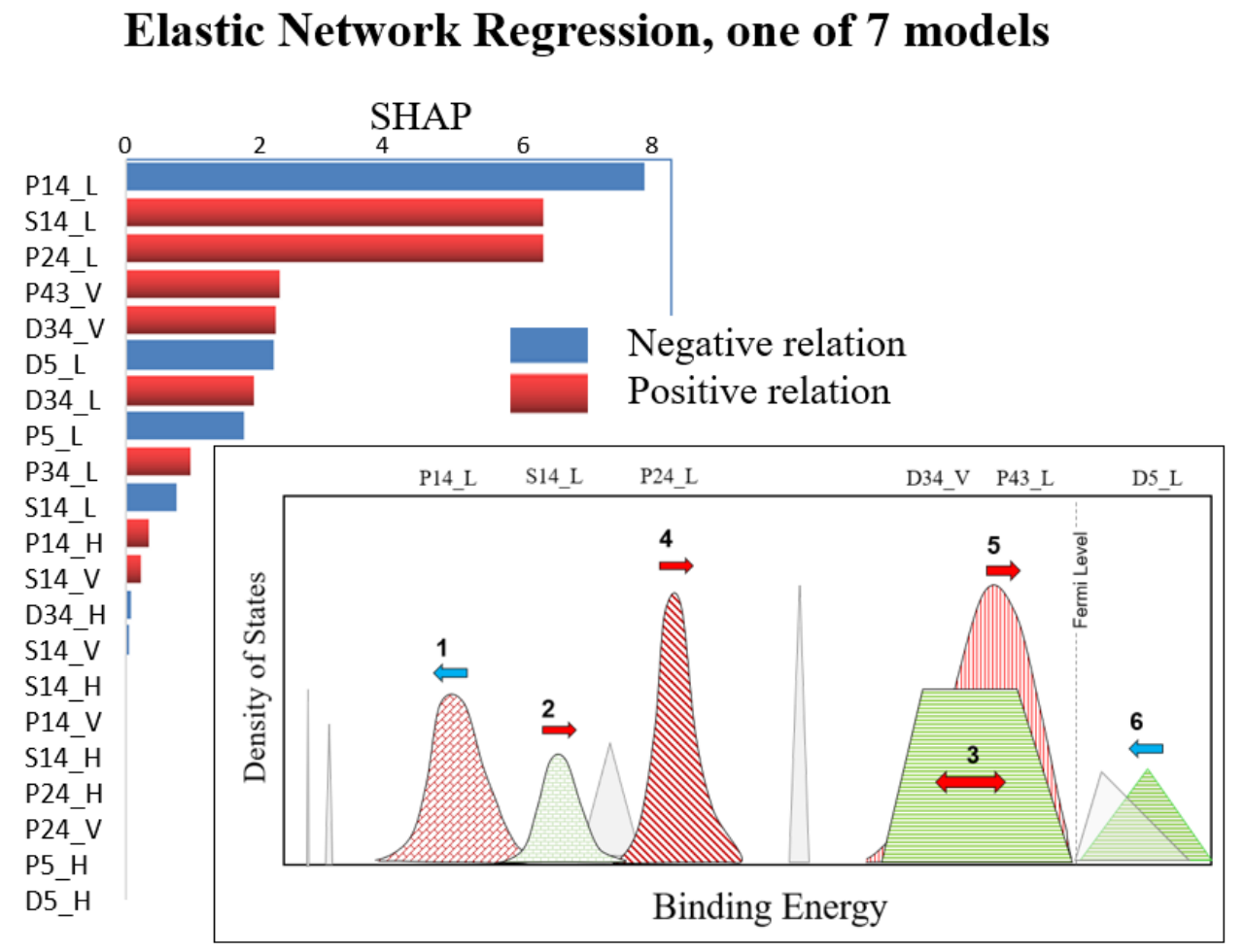

**Fig. 2 Elastic network regression model SHAP analysis (a) bar graph (upper left); (b) a schematic diagram of DoS (right);**

By comparing of the best experimental results of these seven models, it was found that the SHAP weight analysis of 7 algorithms are sometime different in orders, but actually the features ranked in the top 8 were roughly similar for all 7 methods. Among three kinds of peak parameters, the position was the most critical influencing factors, and only two peak half-width factors (D43_W and P34_W) was obtained, but no peak height features were found to influence on $T_c$ maximum, as shown in Table 1.

As shown in Table 1, among different algorithms or models, the SHAP analysis histogram of Bayesian ridge regression model shows a relatively a smaller number of features, with the stronger correlation among P14_L, P24_L, D34_W, and D34_L. From the results of elastic network regression model, the SHAP values of P14_L, S14_L and P24_L are relatively large. For the gradient boost regression model SHAP analysis, the SHAP value of P14_L is unusually large and also has a negative correlation with $T_c$ maximum. The SHAP values of P14_L and S14_L are relatively large in the SHAP analysis of the K-nearest neighbor regression model results. Furthermore, the Lasso regression model is somewhat special, the SHAP values of D34_W, D34_L, P24_L, P14_L are relatively are larger. For the random forest regression model, it shows that the SHAP values of P14_L, D34_W, and P34_L are relatively larger, but with a larger error. Finally, from the support vector machine regression model, results show that the three features of P24_L, P14_L, and S14_L are more strongly correlated with $T_c$ maximum, and also with a larger error.

**Table 1 SHAP weight analysis table and Errors for seven models**

| Model Name | MAE | MSE | Test Index | P14_L | P24_L | D34_W | S14_L | D34_L | D5_L |
|---|---|---|---|---|---|---|---|---|---|
| Bayesian ridge regression[12] | 0.057 | 0.003 | 25 | -7.4 | 10.8 | 7.2 | 2.9 | 6.5 | 5.9 |
| Elastic network regression[13] | 0.138 | 0.019 | 19 | -7.6 | 6.1 | 2.2 | 6.1 | 1.9 | 2.2 |
| Gradient boost regression[14] | 0.123 | 0.015 | 21 | -19.5 | 0.2 | 2.8 | 0.2 | 1.5 | 0.4 |
| K-nearest neighbor regression[15] | 0.167 | 0.028 | 13 | -12.9 | 1.8 | 0.4 | 10.4 | 0.6 | 1.6 |
| Lasso regression[16] | 0.052 | 0.003 | 0 | -5.6 | 7.8 | 12.3 | 2.6 | 8.1 | 4.5 |
| Random forest regression[17] | 0.308 | 0.095 | 19 | -8.8 | 1.4 | 6.4 | 0.4 | 2.2 | 0.6 |
| Support vector machine regression[18] | 0.394 | 0.155 | 20 | -5.6 | 11.1 | 2.5 | 0.1 | 1.2 | 3.1 |
| 7 model mean value | - | - | - | -9.6 | 5.6 | 4.8 | 3.3 | 3.1 | 2.6 |

In summary, although differences between the models, the range of main candidate factors are similar but only with some difference from the SHAP value weight ranks. By compared with each other's, only the random forest and gradient boosting models have a larger error on SHAP ranks, seems these two methods may require a larger amount of data as they usually analyze the problem of large sample data set.

## Discussion

From the governing characteristic parameters of seven ML training results, it was showed that there are two important energy-band features highly influencing on *T*c maximum: the first one is the outer layer coupling energy band from P34 and D43, which is the well-known *pd* coupling around the Fermi energy level, corresponding to the traditional valence band from the copper-oxygen plane. Actually, it is an important law of the flat band feature of cuprate superconductor, which were thought to explain the formal doping law that more suitable conductive carries enhance $T_c$ maximum. Another one is a new inner orbital coupling with P14 and S14, which is a special coupling between 2*s* core orbital of O ion and the *p* core orbital from the cations neighboring cooper-oxygen plane, since two orbitals are similar in energy and space relatively. From all 7 algorithms' results, obviously the second law seems to be more important for higher $T_c$ maximum, which shows that the deeper its position the better *T*c maximum. Obviously, this result is more interesting than that of the formal flat band law, which is the only result obtained from the formal ML of 1,2000 big data set. Obviously, the new result is mined successfully because of the improvement both from data set of *T*c maximum and from selection of a wider range of orbital interaction related with the coper-oxygen plane and the plane's neighbor ions.

In summary, considering on the complex selection of orbitals which might influence the results of ML training tests; therefore, we work should rely on the crystal structure laws of cuprate superconducting, and also should ignored few unregular and unimportant independent orbitals (*i.e.*, introducing domain knowledge to reduce studying factors). Although among these algorithms, some SHAP weight rank order was influenced, and some errors of models were different from each other, which is limited influence on the final conclusion about the key influence factors on *T*c maximum. Actually, no further prediction model was discussed in this work since here we have no idea about any new cuprate superconductor system.

As a result of multiple factors had been obtained by ML, obviously it is hard to do such a work by common statistical method. But if we could find some parameters, directly related with *T*c maximum rather than orbital feathers, we might also search for a simpler statistical law of *T*c maximum, for example the energy span of two energy band was a suitable feature of band structure as our precious work achieved.[9]

As the core orbital coupling found by ML is more important for *T*c maximum than the formal flat band character, we thought that this new direction is reasonable since it quite agrees with Anderson's "More is Different", which was thought to be a basic law of condensed mater physics. Furthermore, according to Moriya-Ueda theory, the frequency spread of the wavevector-dependent part of the spin fluctuation, which is mainly restricted to a bandwidth.[19] Therefore, a bandwidth should be a governing factor to answer the question about the difference of *T*c maximum for each cuprate system.[20] But, it should be known that the "real" bandwidth could be taken as not only *pd* coupling band, but also the new *sp* interaction, in which these two coupling band could be, through 2*s* and 2*p* orbitals of O ion, connected by corelated or entangled with each other orbitals. Therefore, two orbital feature laws should be used to explain the strength of pairing factor, and new idea might help us find some new high-*T*c superconductors, rather than only some suitable doping systems.

# Appendix A Table of the main orbital characteristic parameters obtained from the density of states calculated from DFT method of first principle theory[9]

| Sample name | S14_L | S14_H | S14_V | P14_L | P14_H | P14_V | S14_L | S14_H | S14_V | P24_L | P24_H |
|---|---|---|---|---|---|---|---|---|---|---|---|
| $Bi_2Sr_2C_{a2}Cu_3O_{10}$ | -29.17 | 0.58 | 0.22 | -20.22 | 1.06 | 0.36 | -17.41 | 0.71 | 0.9 | -12.75 | 1.14 |
| $Bi_2Sr_2Ca_3Cu_4O_{12}$ | -34.19 | 0.71 | 0.26 | -21.82 | 1.59 | 0.52 | -20.24 | 0.8 | 1.38 | -16.71 | 0.73 |
| $Bi_2Sr_2CaCu_2O_8$ | -32.58 | 0.59 | 0.26 | -19.87 | 0.59 | 0.39 | -17.6 | 0.91 | 0.68 | -14.43 | 1.13 |
| $Bi_2Sr_2CuO_6$ | -33.9 | 0.93 | 0.13 | -18.42 | 0.97 | 0.19 | -15.88 | 1.75 | 0.1 | -18.42 | 0.97 |
| $DyBa_2Cu_3O_7$ | -25.32 | 0.79 | 0.27 | -22.61 | 0.39 | 0.83 | -17.26 | 1.02 | 0.74 | -10.68 | 1.43 |
| $ErBa_2Cu_3O_7$ | -25.26 | 0.77 | 0.27 | -24.38 | 0.49 | 0.64 | -17.41 | 1.03 | 0.71 | -10.62 | 1.37 |
| $EuBa_2Cu_3O_7$ | -25.15 | 0.73 | 0.27 | -18.68 | 0.3 | 0.99 | -16.39 | 0.8 | 0.85 | -10.47 | 1.35 |
| $GdBa_2Cu_3O_7$ | -25.29 | 0.75 | 0.27 | -19.41 | 0.42 | 0.73 | -17.17 | 0.79 | 0.9 | -10.66 | 1.34 |
| $HgBa_2Ca_2Cu_3O_9$ | -26.73 | 0.68 | 0.26 | -19.74 | 1.41 | 0.38 | -17.08 | 1.01 | 0.8 | -11.96 | 1.39 |
| $HgBa_2Ca_3Cu_4O_{11}$ | -24.81 | 0.48 | 0.26 | -19.67 | 1.47 | 0.39 | -16.94 | 1.25 | 0.55 | -10.05 | 0.88 |
| $HgBa_2CaCu_2O_7$ | -26.67 | 0.93 | 0.26 | -19.74 | 1.06 | 0.35 | -17.15 | 0.92 | 0.93 | -11.89 | 1.91 |
| $HoBa_2Cu_3O_7$ | -25.23 | 0.78 | 0.27 | -22.02 | 0.52 | 0.61 | -17.36 | 1.05 | 0.71 | -10.61 | 1.39 |
| $La_{1.85}Ba_{0.15}CuO_4$ | -24.95 | 0.11 | 1.03 | -15.99 | 2.04 | 0.16 | -18.22 | 0.89 | 0.26 | -10.35 | 0.26 |
| $La_{1.85}Sr_{0.15}CuO_4$ | -32.53 | 1.37 | 0.08 | -15.97 | 2.03 | 0.08 | -18.2 | 0.86 | 0.27 | -15.97 | 2.03 |
| $LaBa_2Cu_3O_7$ | -25.26 | 0.72 | 0.26 | -14.56 | 0.47 | 0.59 | -17.27 | 0.67 | 0.99 | -10.57 | 1.37 |
| $La_2CuO_4$ | -32.58 | 1.57 | 0.13 | -16.03 | 2.21 | 0.14 | -18.25 | 0.9 | 0.92 | -16.03 | 2.21 |
| $LuBa_2Cu_3O_7$ | -25.34 | 0.76 | 0.28 | -25.42 | 1.16 | 0.27 | -17.46 | 0.91 | 0.81 | -10.74 | 1.34 |
| $Nd_{1.32}Ce_{0.4}1Sr_{0.27}CuO_4$ | -33.08 | 0.53 | 0.41 | -15.35 | 0.85 | 0.38 | -17.28 | 0.55 | 0.78 | -15.35 | 0.85 |
| $Nd_{1.85}Ce_{0.15}CuO_4$ | -33.32 | 0.69 | 0.3 | -15.12 | 0.93 | 0.34 | -18.72 | 0.57 | 0.73 | -15.12 | 0.93 |
| $NdBa_2Cu_3O_7$ | -25.32 | 0.73 | 0.27 | -19 | 0.27 | 1.09 | -16.43 | 0.74 | 0.93 | -10.66 | 1.36 |
| $PrBa_2Cu_3O_7$ | -25.39 | 0.77 | 0.26 | -18.67 | 0.31 | 1.01 | -16.36 | 0.66 | 1.08 | -10.71 | 1.53 |
| $SmBa_2Cu_3O_7$ | -25.24 | 0.73 | 0.27 | -18.72 | 0.28 | 1.04 | -16.51 | 0.74 | 0.92 | -10.54 | 1.35 |
| $Sr_{0.7}Ca_{0.3}CuO_2$ | -32.73 | 0.76 | 0.05 | -20.4 | 1.01 | 0.03 | -17.13 | 0.85 | 0.09 | -14.64 | 1.44 |
| $Sr_{0.9}La_{0.1}CuO_2$ | -33.21 | 0.89 | 0.09 | -15.21 | 1.74 | 0.07 | -17.88 | 0.74 | 0.22 | -15.22 | 1.74 |
| $T_{l2}Ba_2Ca_2Cu_3O_{10}$ | -26.67 | 0.52 | 0.27 | -19.73 | 1.09 | 0.38 | -17.09 | 0.78 | 0.88 | -11.73 | 0.91 |
| $Tl_2Ba_2Ca_3Cu_4O_{12}$ | -24.63 | 0.44 | 0.27 | -19.76 | 1.33 | 0.4 | -17 | 0.96 | 0.75 | -9.87 | 0.81 |
| $Tl_2Ba_2CaCu_2O_8$ | -26.58 | 0.81 | 0.26 | -19.51 | 0.88 | 0.36 | -18.63 | 0.95 | 0.9 | -11.98 | 1.4 |
| $Tl_2Ba_2CuO_6$ | -26.22 | 0.8 | 0.26 | -11.3 | 1.31 | 0.24 | -18.48 | 0.76 | 0.83 | -11.3 | 1.31 |
| $TlBa_2Ca_2Cu_3O_9$ | -24.61 | 0.53 | 0.27 | -19.72 | 1.14 | 0.38 | -16.99 | 0.83 | 0.79 | -9.88 | 0.98 |
| $TlBa_2Ca_3Cu_4O_{11}$ | -24.61 | 0.45 | 0.27 | -19.68 | 1.3 | 0.42 | -17.02 | 0.93 | 0.71 | -9.86 | 0.82 |
| $TlBa_2CaCu_2O_7$ | -24.68 | 0.66 | 0.27 | -19.7 | 0.84 | 0.32 | -16.25 | 0.86 | 0.72 | -9.92 | 1.21 |
| $TlBaSrCuO_5$ | -23.98 | 0.49 | 0.27 | -14.03 | 1.24 | 0.32 | -16.73 | 0.75 | 0.89 | -9.19 | 0.88 |
| $TmBa_2Cu_3O_7$ | -25.2 | 0.77 | 0.27 | -23.72 | 0.49 | 0.65 | -17.34 | 1 | 0.74 | -10.61 | 1.34 |
| $YBa_2Cu_3O_7$ | -25.27 | 0.78 | 0.27 | -21.61 | 0.88 | 0.35 | -17.31 | 1.06 | 0.69 | -10.64 | 1.37 |
| $YbBa_2Cu_3O_7$ | -25.21 | 0.74 | 0.27 | -22.51 | 0.7 | 0.43 | -17.22 | 1.02 | 0.69 | -10.59 | 1.31 |

| Sample name | P24_V | P43_L | P43_V | D34_L | D34_H | D34_V | P5_L | P5_H | D5_L | D5_H | Tc/K |
|---|---|---|---|---|---|---|---|---|---|---|---|
| **$Bi_2Sr_2Ca_2Cu_3O_{10}$** | 0.17 | -2.75 | 1.91 | -2.29 | 0.5 | 1.71 | 3.52 | 0.39 | 3.55 | 0.04 | 110 |
| **$Bi_2Sr_2Ca_3Cu_4O_{12}$** | 0.38 | -5.16 | 3.33 | -4.99 | 0.62 | 2.7 | 1.38 | 0.65 | 1.31 | 0.14 | 110 |
| **$Bi_2Sr_2CaCu_2O_8$** | 0.41 | -2.77 | 1.86 | -2.53 | 0.46 | 1.53 | 2.18 | 0.54 | 1.9 | 0.03 | 96 |
| **$Bi_2Sr_2CuO_6$** | 0.19 | -2.09 | 1.09 | -2.68 | 0.32 | 0.86 | 1.14 | 0.28 | 1.03 | 0.04 | 34 |
| **$DyBa_2Cu_3O_7$** | 0.45 | -2.28 | 2.26 | -2.86 | 0.98 | 1.48 | 4.13 | 0.19 | 5.13 | 0.25 | 95.1 |
| **$ErBa_2Cu_3O_7$** | 0.46 | -2.82 | 2.2 | -2.86 | 0.95 | 1.49 | 4.08 | 0.2 | 4.13 | 0.18 | 92.4 |
| **$EuBa_2Cu_3O_7$** | 0.43 | -2.16 | 2.05 | -2.15 | 0.78 | 1.69 | 4.07 | 0.18 | 5.93 | 0.21 | 95 |
| **$GdBa_2Cu_3O_7$** | 0.45 | -2.29 | 2.13 | -2.88 | 0.93 | 1.47 | 3.91 | 0.21 | 5.42 | 0.28 | 94.5 |
| **$HgBa_2Ca_2Cu_3O_9$** | 0.39 | -2.65 | 2.41 | -1.84 | 0.74 | 1.62 | 0.78 | 0.27 | 4.25 | 0.18 | 135 |
| **$HgBa_2Ca_3Cu_4O_{11}$** | 0.43 | -2.36 | 2.07 | -1.87 | 0.7 | 1.62 | 4.42 | 0.19 | 4.47 | 0.06 | 125 |
| **$HgBa_2CaCu_2O_7$** | 0.38 | -2.6 | 2.57 | -2.21 | 0.74 | 1.49 | 4.08 | 0.24 | 4.23 | 0.42 | 128 |
| **$HoBa_2Cu_3O_7$** | 0.46 | -2.79 | 2.22 | -2.88 | 0.98 | 1.46 | 4.02 | 0.22 | 4.13 | 0.18 | 92.9 |
| **$La_{1.85}Ba_{0.15}CuO_4$** | 0.1 | -3.81 | 0.69 | -2.61 | 0.76 | 0.34 | 3.4 | 0.26 | 3.41 | 0.05 | 43 |
| **$La_{1.85}Sr_{0.15}CuO_4$** | 0.08 | -3.8 | 0.69 | -2.62 | 0.77 | 0.33 | 3.41 | 0.26 | 3.23 | 0.04 | 36 |
| **$LaBa_2Cu_3O_7$** | 0.41 | -2.11 | 1.97 | -2.79 | 0.92 | 1.37 | 3.89 | 0.17 | 5.55 | 0.23 | 93 |
| **$La_2CuO_4$** | 0.14 | -4.42 | 2.49 | -2.54 | 0.74 | 1.26 | 3.78 | 0.22 | 4.99 | 0.85 | 41.5 |
| **$LuBa_2Cu_3O_7$** | 0.47 | -2.94 | 2.21 | -2.92 | 0.95 | 1.49 | 4.07 | 0.19 | 5.33 | 0.33 | 92.4 |
| **$Nd_{1.32}Ce_{0.4}1Sr_{0.27}CuO_4$** | 0.38 | -2.92 | 1.29 | -3.14 | 0.73 | 0.66 | 0.77 | 0.2 | 0.79 | 0.09 | 24 |
| **$Nd_{1.85}Ce_{0.15}CuO_4$** | 0.34 | -3.05 | 1.25 | -3.18 | 0.7 | 0.67 | 0.64 | 0.2 | 0.65 | 0.09 | 24 |
| **$NdBa_2Cu_3O_7$** | 0.43 | -2.16 | 2.05 | -2.83 | 0.94 | 1.4 | 3.88 | 0.21 | 3.87 | 0.16 | 96 |
| **$PrBa_2Cu_3O_7$** | 0.4 | -2.12 | 2.15 | -2.78 | 1 | 1.39 | 4.11 | 0.21 | 3.96 | 0.18 | 92 |
| **$SmBa_2Cu_3O_7$** | 0.43 | -2.1 | 2.04 | -2.86 | 0.93 | 1.42 | 4.02 | 0.18 | 5.69 | 0.16 | 93.5 |
| **$Sr_{0.7}Ca_{0.3}CuO_2$** | 0.05 | -2.49 | 0.22 | -1.72 | 0.81 | 0.2 | 1.91 | 0.14 | 1.02 | 0.1 | 99 |
| **$Sr_{0.9}La_{0.1}CuO_2$** | 0.07 | -2.92 | 0.49 | -2.1 | 0.94 | 0.39 | 3.53 | 0.36 | 4.43 | 0.05 | 43 |
| **$Tl_2Ba_2Ca_2Cu_3O_{10}$** | 0.45 | -2.15 | 2.06 | -1.72 | 0.55 | 1.69 | 2.92 | 0.18 | 3.21 | 0.09 | 128 |
| **$Tl_2Ba_2Ca_3Cu_4O_{12}$** | 0.44 | -2.48 | 2.14 | -2.43 | 0.67 | 1.6 | 3.35 | 0.18 | 3.54 | 0.03 | 119 |
| **$Tl_2Ba_2CaCu_2O_8$** | 0.46 | -2.56 | 2.56 | -2.85 | 0.58 | 1.65 | 2.92 | 0.23 | 3.39 | 0.19 | 110 |
| **$Tl_2Ba_2CuO_6$** | 0.24 | -1.91 | 1.9 | -2.83 | 0.35 | 1.35 | 4.17 | 0.21 | 4.59 | 0.19 | 90 |
| **$TlBa_2Ca_2Cu_3O_9$** | 0.44 | -2.64 | 1.94 | -2.38 | 0.67 | 1.46 | 4.43 | 0.34 | 4.55 | 0.08 | 133 |
| **$TlBa_2Ca_3Cu_4O_{11}$** | 0.44 | -2.6 | 1.99 | -2.32 | 0.69 | 1.58 | 4.22 | 0.21 | 4.15 | 0.07 | 127 |
| **$TlBa_2CaCu_2O_7$** | 0.44 | -2.51 | 1.85 | -2.43 | 0.59 | 1.36 | 4.28 | 0.39 | 4.89 | 0.14 | 82 |
| **$TlBaSrCuO_5$** | 0.45 | -2.03 | 2 | -2.43 | 0.4 | 1.48 | 6.64 | 0.43 | 6.72 | 0.21 | 56.5 |
| **$TmBa_2Cu_3O_7$** | 0.47 | -2.85 | 2.21 | -2.91 | 0.94 | 1.51 | 3.82 | 0.22 | 4.95 | 0.26 | 92.5 |
| **$YBa_2Cu_3O_7$** | 0.46 | -2.78 | 2.2 | -2.91 | 0.97 | 1.46 | 3.9 | 0.22 | 4.9 | 0.36 | 92 |
| **$YbBa_2Cu_3O_7$** | 0.46 | -2.78 | 2.11 | -2.91 | 0.97 | 1.4 | 4.11 | 0.18 | 5.89 | 0.23 | 96.1 |

**Note: L: energy location to Fermi level, W: half-peak width, H: relative height. In addition to the commonly S, P and D are used for these orbital such as *s*, *p* and *d* orbital peak, the ions 1# and 2# neighboring cations represent for far away and near Fermi level respectively, as well as Cu ion and O ion in the copper-oxygen plane are named as ions 3# and 4# etc. as the relevant orbital ion position corner scale, L, W and H represents respectively as the position of the relevant peak, half height width and relative, in which the half peak width W is coming from the total number of electrons on the orbital divided by the peak height of peak. And the peak relative height H is taken as the peak height of P34 as 1, and the other peak heights are taken as its relative value.**